\documentclass[runningheads,a4paper]{llncs}

\usepackage{amssymb}
\usepackage{graphicx}
\usepackage[misc,geometry]{ifsym} 
\usepackage[hyphens]{url}

\usepackage{url}
\urldef{\mailsa}\path|mireia.yurrita@gmail.com|    
\newcommand{\keywords}[1]{\par\addvspace\baselineskip
\noindent\keywordname\enspace\ignorespaces#1}
\usepackage{hyperref}
\usepackage{cleveref}
\usepackage{microtype}
\hypersetup{
    colorlinks = true,
    linkcolor = blue,
    filecolor = blue,
    urlcolor = blue,
    citecolor = blue,
}
    
\begin{document}

\mainmatter 

\title{Dynamic Urban Planning: an Agent-Based Model Coupling Mobility Mode and Housing Choice. Use case Kendall Square. }

\titlerunning{Dynamic Urban Planning}

\author{Mireia Yurrita  \textsuperscript{(\Letter)}%
\and Arnaud Grignard \and Luis Alonso \and Yan Zhang \and\\
 Cristian Jara-Figueroa\and Markus Elkatsha \and Kent Larson}

\authorrunning{M. Yurrita et al.}

\institute{Massachusetts Institute of Technology, Cambridge, USA \\
\mailsa\\  
} 

\toctitle{Lecture Notes in Computer Science}
\tocauthor{Authors' Instructions}
\maketitle

\begin{abstract}
As cities become increasingly populated, urban planning plays a key role in ensuring the equitable and inclusive development of metropolitan areas. MIT City Science group created a data-driven tangible platform, CityScope, to help different stakeholders, such as government representatives, urban planners, developers, and citizens, collaboratively shape the urban scenario through the real-time impact analysis of different urban interventions. This paper presents an agent-based model that characterizes citizens' behavioural patterns with respect to housing and mobility choice that will constitute the first step in the development of a dynamic incentive system for an open interactive governance process. The realistic identification and representation of the criteria that affect this decision-making process will help understand and evaluate the impacts of potential housing incentives that aim to promote urban characteristics such as equality, diversity, walkability, and efficiency. The calibration and validation of the model have been performed in a well-known geographic area for the Group:  Kendall Square in Cambridge, MA.

\end{abstract}

\section{Introduction}

Even if humans started to cluster around cities 5,500 years ago, it was not until about 150 years ago that cities became increasingly populated \cite{CitiesEvolution}. Nowadays more than 50\% of the world's population lives in urban areas, which makes urban planning emerge as one of the key elements for citizens' well-being \cite{CityScope}. As in the following decades cities may account for 60\% of the total global population, 75\% of the global $C0_2$ emissions, and up to 80\% of all world energy use \cite{UNReport}, traditional systems of centralised planning might be obsolete \cite{CityScope}. 

\begin{figure}[]
    \centering
    \includegraphics[scale = 0.25]{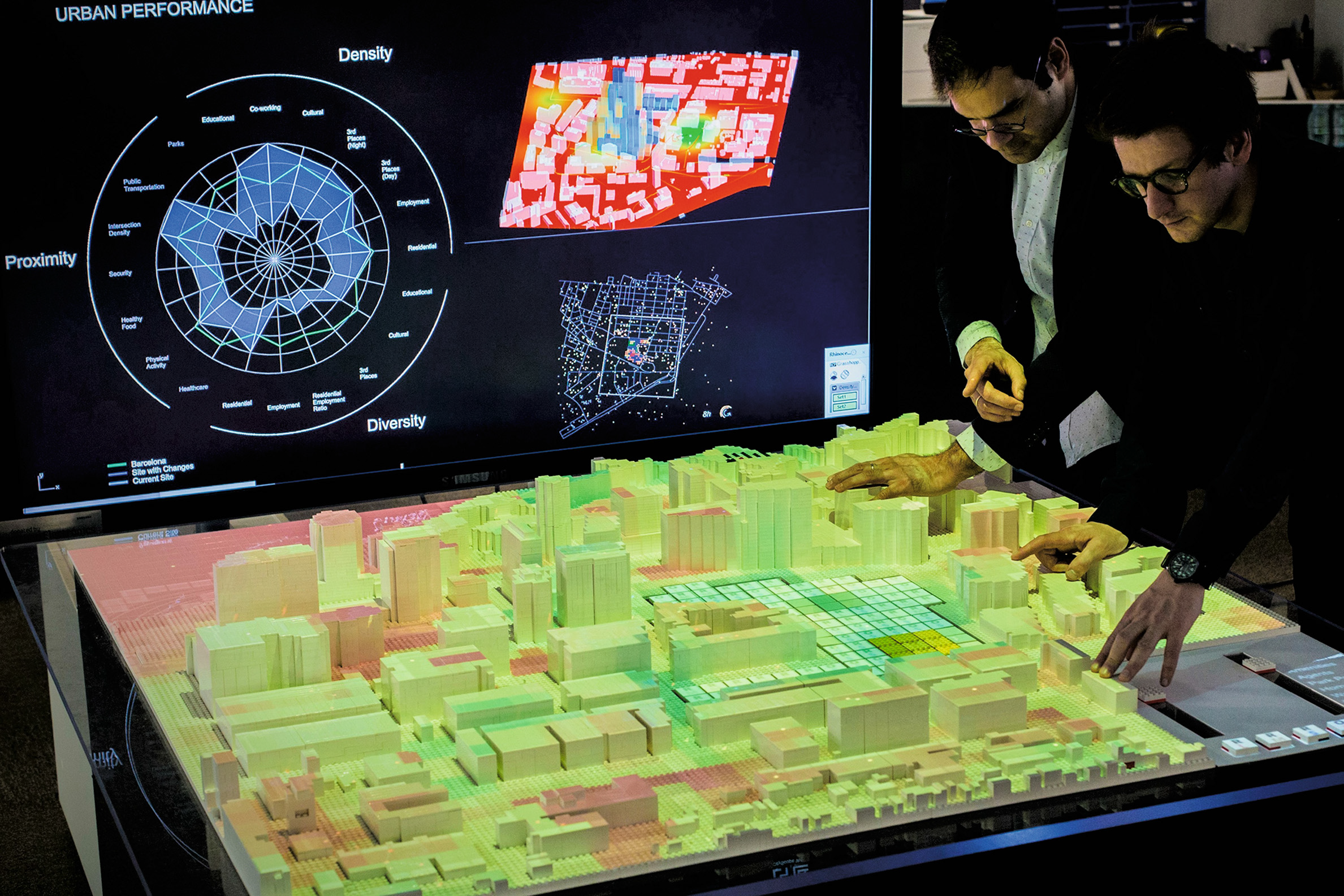}
    \caption{CityScope Volpe: a data-driven interactive simulation tool for urban design  \cite{CityScope}}
    \label{fig:cityScope}
\end{figure}

In view of the need for more participatory decision-making processes, MIT City Science (CS) Group developed a tangible data-driven platform, CityScope. This platform brings different stakeholders together in an effort to enable a more collaborative urban design process \cite{CityScope} \cite{CityScope2} --Figure \ref{fig:cityScope}--. As part of this collective decision-making approach, the CS Group is designing a dynamic incentive system, that aims to promote pro-social urban features through dynamic incentive policies. By using data-driven agent-based simulations, the impacts on equality, diversity, walkability, and efficiency of such incentives are measured. Thanks to this information, different stakeholders, such as government representatives, urban planners, developers, and citizens, can collaboratively decide which intervention shapes the most favourable urban scenario. 

Some of the incentives explored in this line of research are focused on the promotion of affordable housing within a 20-minute walking distance from one's workplace. This vision is part of a broader perspective where cities are encouraged to be comprised of autonomous communities, where citizens' daily needs are met within a walking distance from their doorstep. MIT City Science intends to create the technological tools to facilitate consensus between community members, local authorities, and design professionals \cite{CityScope}, so that cities can fulfil their ambition of having walkable districts.

This paper arises from the need to forecast citizens' criteria when choosing their homes, and the consequent mobility patterns. By doing so the effects of numerous pro-social incentives that aim to address the deficiency of affordable housing in city centres can be tested. Although some previous studies focusing on citizens' mobility choices had already been held in the group, the need to combine mobility trends with housing-related aspects was major. In contrast to the consulted bibliography, where most approaches make use of statistical tools to respond to this joint choice research question, Agent-Based Modeling (ABM) has been deployed in this study. People's behavioral patterns have been represented based on their income profiles.

This document is organized as follows. Section \ref{StateOfArt} describes previous work on the representation of housing and residential mobility choices and highlights the benefits of using ABM when embracing the complexity of urban dynamics. Section \ref{model} gives a detailed summary of the agents that constitute the model, the behaviours of each of them and the dynamics of the housing and mobility mode choice. Section \ref{useCase} calibrates the model for the particular use case of Kendall Square using census and transportation data and section \ref{conclusion} gives some final thoughts on the effectiveness of the method. 

\section{Background} \label{StateOfArt}

\subsection{Related Work}

The present research is an agent-based joint representation of housing and mobility choices that intends to forecast citizens' behavioural patterns when choosing their residencies as well as to highlight impacts on mobility. This study has been built on top of previous work where different approaches to address this question have been suggested. Some of them merely focused on the housing choice search, others concentrated on the mobility mode issue and some merged both points of view.

Among the former methodologies, a two-stage behavioural housing search model \cite{twoStage} had been created. This model was comprised of a hazard-based ``choice set formation step'', and a ``final residential location selection step'' using a multinomial logit formulation. A theoretical housing preference and choice framework had also been given using the theory of mean-end chain \cite{housingChoiceTheory}. The motivations that make people want to move had been studied \cite{intentionMove} using a nested logit model, which paid special attention to how residential decisions are impacted by transportation. A multi-agent approach had also been used to model and analyze the residential segregation phenomenon \cite{residentialSegregation}. And an ABM had been designed to represent the dynamics of the housing market and for the study of the effects that certain parameters, like school accessibility, would have on the outcome \cite{ABMHousing}.

As far as mobility choices are concerned, some previous studies had been carried out within the MIT City Science Group using both statistical tools as well as agent-based models. The HCl platform deployed for the real-time prediction of mobility choices through discrete choice models \cite{MoCho} is one of the examples of how the CS Group had previously addressed the mobility choice issue. Agent-based models had also been developed in our Group to evaluate new mobility mode choices and to assess their impact on cities \cite{GameIt}.

Among previous joint choice models, it is important to highlight a model based on the disaggregate theory where the combinations of location, housing, automobile ownership and the commuting mode had been studied \cite{jointChoiceAutomobile}. Additionally, a nested multinomial logit model to estimate the joint choice of residential mobility, and housing choice \cite{empiricalTest} had also been used.

This paper presents a novel joint choice model based on agents. Following the Schelling segregation model \cite{Schelling}, the factors related to housing, and those related to transportation will be deployed to determine if a certain agent is willing to move or not.

\subsection{Agent-Based Simulations}

Most of the related work mentioned prior in this paper make use of statistical tools to study the housing and residential mobility choice phenomenon. Our research, on the contrary, will artificially reproduce the dynamics of the urban scenario using agent-based simulations. Although several platforms dedicated to agent-based modelling have been created lately, this model has been developed using the GAMA platform. GAMA allows modellers to create complex multi-level spatially explicit simulations where GIS data can be easily integrated. It also provides a high-level agent-oriented language, GAML, as well as an integrated development environment \cite{GAMA}\cite{GAMA2}. The behaviour of each `people agent' has been developed using this tool and some agent-specific functions have been designed in order to recreate their decision-making reasoning while representing the complexity of the surrounding dynamics.

\section{Model description} \label{model}

This model aims to represent the criteria adopted by citizens when choosing their residential location and mobility mode, as well as the importance given to each of these parameters. The realistic characterization of these behaviours enables the study of the consequences that certain housing incentives and urban disruptions might entail.

\subsection{Entities, state variables, and scales}

The \textbf{environmental variables} used to shape the urban scenario and represent its dynamics include:
\begin{itemize}
    \item \textbf{Census block group:} polygon representing the combination of census blocks, smallest geographic areas for which the US Bureau of Census collects data \cite{USGovData}. Each census block group will be formed by the following attributes: \textit{GEOID}: unique geographical ID for each census block group, \textit{vacant\_spaces}: number of vacant housing options available within the boundaries of this polygon (previously a web scraping process has been held and the availability of accommodation and its price identified \cite{Padmapper} within the area of interest), \textit{city}: broader unit the block group belongs to \cite{USGovData}, \textit{rent\_vacancy}: mean rent for each housing option based on online rent information \cite{Padmapper}, \textit{population}: map indicating the number of citizens of each income profile that actually live in the area \cite{USCensus}, \textit{has\_T}: boolean attribute pointing out whether the block group has a T station or not \cite{MBTA} and \textit{has\_bus}: boolean attribute that indicates if bus services are available in the area \cite{MBTA}. Census block groups have been the environmental variables chosen to represent the broadest granularity in the model, since applying the same unit as the US Census Bureau enables the calibration and validation of the model using census data.
    \item \textbf{Building:} polygon representing a finer granularity for the area where the urban interventions are considered to take place. The aforementioned census block groups are the alternative to living in a building within the area of interest. The attributes of `building' agents include: \textit{associated\_block\_group}: block group in which the building is located \cite{USGovData}, \textit{vacant\_spaces}: number of vacant dwelling units within the building \cite{Padmapper} and \textit{rent\_vacancy}: mean rent for each housing option within the building \cite{Padmapper}.
    \item \textbf{Road}: network of roads that agents can use to move around. \textit{mobility\_allowed} is the only attribute of the road network, which is taken into account when considering different commuting mobility options and when the resulting time for each of them is calculated \cite{USGovData}.
\end{itemize}

\begin{figure}[]
    \centering
    \includegraphics[scale = 0.45]{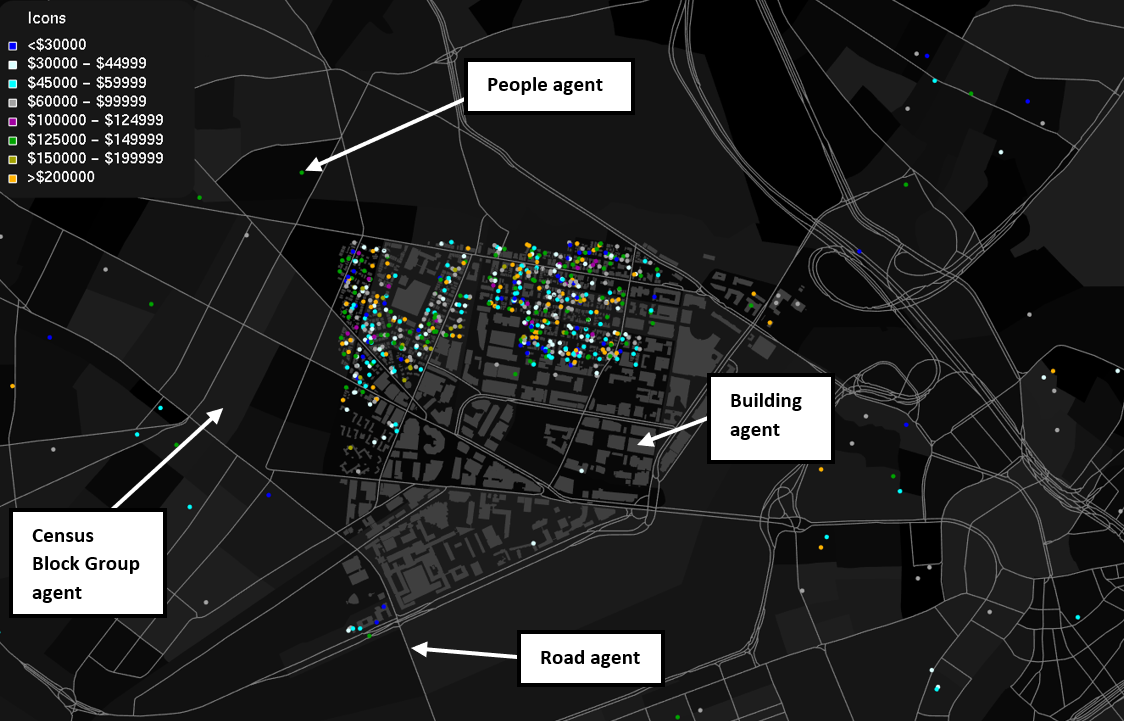}
    \caption{Model overview, where People, Building, Road and Census Block Group --areas with different transparency-- agents have been illustrated for Kendall Square in Cambridge, MA.}
    \label{fig:modelOverview}
\end{figure}

As far as \textbf{agents} are concerned, `people agents' are defined the following way:

\begin{itemize}
    \item \textbf{People}: citizens who work in the area of interest and whose objective is to find the most appropriate housing and mobility option based on their criteria. Their attributes include: \textit{type}: type of `people agent' based on their annual income profile \cite{USCensus}, \textit{living\_place}: building or census block group where they decide to live in each iteration of the searching process, \textit{activity\_place}: building in the area of interest where each agent works, \textit{possible\_mobility\_modes}: list of mobility modes that are available for each agent considering if they own some kind of private mobility mode \cite{USCensus} or if in their living place there is public transportation available \cite{MBTA}, \textit{mobility\_mode} commuting mobility mode chosen in each iteration of the housing search, \textit{time\_main\_activity}: commuting time in minutes, \textit{distance\_main\_activity}: commuting distance depending on the housing option being considered and \textit{commuting\_cost}: cost based on the distance and the mobility mode chosen.
\end{itemize}

\subsection{Process overview}

The dynamics of this model are ruled by the behaviour of `people agents'. Each of them is assigned a nonresidential building in the finer grained area as their workplace. An iterative process will be performed in order to find the housing option and mobility mode that best meets their needs. 
\subsubsection{Iteration 0:}
The initialisation process takes place in three different steps: (i) a random residential unit is assigned to each agent, (ii) based on the area where this residential unit is located and the probability of owning private means of transportation, possible mobility modes are defined (iii) travelling options will be weighed based on each profile's criteria and a final score for each mode will be calculated as a linear combination of the importance given to each of these factors. The option that maximizes the score will be the chosen mobility mode. Criteria related to transportation preferences include both quantitative and qualitative factors such as price, resulting commuting time, the difficulty of usage and the social pattern \cite{GameIt}. Table \ref{tab:transpWeights} displays some synthetic values of this criteria for three medium income profiles \cite{GameIt} (modified from working status profiles into income profiles), whereas table \ref{tab:transpFeatures} shows the features of some of the available transportation modes \cite{GameIt} \cite{MBTA}. \\

\begin{table}[]
\centering
\begin{tabular}{lllll}\hline
                  & Price [-] & Time [-]  & Difficulty [-] & Pattern [-] \\\hline
\$45,000-\$59,999   & -0.8  & -0.8  & -0.7       & 0.7            \\\hline
\$60,000-\$99,999   & -0.7  & -0.85 & -0.75      & 0.8            \\\hline
\$100,000-\$124,999 & -0.6  & -0.9  & -0.8       & 0.95      \\\hline \\
\end{tabular}
\caption{Synthetic weighing values given to transportation criteria \cite{GameIt} according to different income profiles. These values are normalized between -1 and 0 for price, time and difficulty and between 0 and 1 for pattern importance. They will determine each `people agent's ' preference towards a particular mobility choice.}
\label{tab:transpWeights}
\end{table}

\newpage
Note that, in order to be able to make use of public transportation, both the surroundings of the residential unit and the workplace have to offer that particular commuting option. So as to calculate the time needed to commute for each mode, both the waiting time (when non-private modes are considered) and the commute itself --following the topology of the transportation network-- have been taken into account.\\

\begin{table}[]
\centering
\begin{tabular}{lccc}\hline
     & Price/km {[}\${]} & \begin{tabular}[c]{@{}l@{}}Mean speed {[}km/h{]}\end{tabular} & Pattern [-] \\\hline
bike  & 0.01      & 5                                                              & 0.5     \\\hline
bus & 0.1     & 20                                                               & 0.4     \\\hline
car  & 0.32     & 30                                                              & 1     \\ \hline\\
\end{tabular}
\caption{Price, speed and social pattern features of some of the available mobility modes \cite{GameIt}. Price/km and mean speed values are based on the Massachusetts Bay Transportation Authority data \cite{MBTA}, whereas the pattern weight is a synthetic normalized weighing value between 0 and 1. A linear combination between these values and the mobility choice criteria for each income profile is performed so as to select the most suitable transportation option.}
\label{tab:transpFeatures}
\end{table}

\subsubsection{Subsequent iterations:} Once the initialisation process has been held, the procedure will change into: (i) an alternative housing option is randomly assigned to each agent (ii) this new housing option is compared to the current unit and the suitability of moving to the new alternative is evaluated based on certain factors according to each income profile (iii) the `time' parameter is calculated as the resulting commuting time required using the most suitable mobility option for each housing unit as described for iteration 0. `People agents' will decide to move in iteration `i' if the score of the alternative housing unit is greater than the current one. Factors considered when assessing housing include quantitative data such as price or commuting time using the most suitable available means of transportation and qualitative factors like the zone preference, importance given to the unit being in this zone and diversity acceptance \cite{ABMHousing}.\\

\begin{table}[]
\centering
\begin{tabular}{lccc}\hline
                  & Price [-] & Diversity acceptance [-] & Zone weight [-] \\\hline
\$45,000-\$59,999   & -0.8  & 0                    & 0.4         \\\hline
\$60,000-\$99,999   & -0.6  & -0.3                 & 0.5         \\\hline
\$100,000-\$124,999 & -0.5  & -0.5                 & 0.6    \\\hline    \\
\end{tabular}
\caption{Price, diversity acceptance and zone criteria according to different income profiles (synthetic data based example). Values are normalized between -1 and 0 for price, between -1 and 1 for diversity acceptance and between 0 and 1 for zone weight. These criteria, along with the transportation criteria will point out the housing and mobility choice that maximizes the score obtained through a linear combination.}
\label{tab:homeCriteria}
\end{table}

Diversity is calculated using the Shannon-Weaver formula following the diversity calculation methodology deployed on the CityScope platform \cite{CityScope}.  
This measurement quantitatively measures the amount of species (different income profiles in our research) in an ecosystem (the spatial unity being considered in each case) \cite{CityScope}. This diversity metrics will be applied either to a building, if the housing unit being considered is within the finer grained area, or to a census block group should the dwelling be located on the outskirts. This iterative process will be applied until the number of people moving in each iteration asymptotically approaches zero. Once this situation is reached and every `people agent' is assigned the most convenient mobility and housing option, further mobility studies can be performed departing from this t=0 situation.

\subsection{Initialization and input data}

The model initialization relies on two main types of files: the ones containing geographical information and the ones containing information regarding agents and mobility modes. The first group of files includes (1) a shapefile that consists of the block groups of the area in question,  (2) a shepefile where the availability of apartments according to each block group is displayed, (3) a shapefile where the buildings of the finer grained area are incorporated, (4) a shapefile that is comprised of the public transportation system and (5) a shapefile with the road network of the region. 
The second type of input data consists of \textit{.csv} files where (1) income profiles are defined and their characteristics such as the proportion within the population, the probability of owning a car and a bike are listed \cite{USCensus}, (2) `people' profiles define the weight given to each of the criterion when choosing a house --Table \ref{tab:homeCriteria}--, (3) these same profiles describe the criteria regarding mobility mode preferences --Table \ref{tab:transpWeights}-- and (4) different mobility modes are listed and their characteristics detailed --Table \ref{tab:transpFeatures}--, including the price and time per distance unit, the waiting time, the difficulty, and the social pattern.

\section{Model calibration and validation. Use case Kendall Square} \label{useCase}

The model described above in this paper presents a generic framework where, as long as the input data is modified, the methodology can be applied to any city. In order to develop a first version of the model, Kendall Square has been analysed. Kendall Square is a neighbourhood in Cambridge, Massachusetts, --Figure \ref{fig:modelOverview}-- that has been transformed from an industrial area into a leading biotech and IT company hub in the last decades. As the number of companies clustering around this site has soared, the cost of housing has exploded, making it increasingly difficult for low and medium income members to live within a walking distance from their workplace \cite{HousingProblemsKendall}. The commuting patterns derived from this situation, which include high private vehicle density and saturation of the public transportation infrastructure, lead to serious mobility challenges. This section is focused on the description of the calibration process of `people agents' ' behavioural criteria, so that their reactions to potential housing incentives can be characterized and the consequences of their acts monitored. Input files (2) and (3) gather the aforementioned criteria and will be, therefore, inferred from this validation process, whereas file (1) is deduced from census data \cite{USCensus} and file (4) is an adaptation of the Massachusetts Bay Transportation Authority information \cite{MBTA}.

The calibration process is based on the definition of two types of errors: one related to housing choices and the second one related to mobility mode patterns. Batch experiments are then deployed so as to find the combination of criteria that leads to the minimum value of the sum of both of these errors. There are eight different factors that affect the decision-making process of `people agents' --housing price, diversity acceptance, preferred zone and weight given to zone \cite{ABMHousing};  transportation price, time importance, commuting difficulty and social pattern \cite{GameIt}-- which applied to eight different income profiles, leads to a total of sixty four variables handled by batch experiments.

The housing error is defined as the difference in percentage of the distribution of each income profile in each census block group where citizens working in Kendall have certain presence. The real percentage value inferred from transportation data  --the census block group where different Kendall workers live can be identified and the income profile of these approximated to the mean annual income based on census data \cite{USCensus}-- is then compared to the resulting simulated value and the root-mean-square deviation calculated --Equation \ref{eq:RMSE}--. 

\begin{equation} \label{eq:RMSE}
    RMSE = \sqrt{\frac{1}{n} \sum_{i=1}^{n} (Y_i  - \hat{Y}_i)^2} 
\end{equation}

$Y_i$ = real percentage of people of type i living in the neighbourhood in question

$\hat{Y}_i$ = percentage of `people agents' of type i living in that same neighbourhood as a result of the housing selection methodology\\

As for the mobility mode error, the resulting simulated percentages of pedestrians and car, bike, bus and T usage are compared to the Parking and Transportation Demand Management Data in the City of Cambridge \cite{TransportCambridge}. This error is equally calculated as the root-mean-square deviation, where $Y_i$ represents the percentage of people that choose a certain mobility mode i, whereas $\hat{Y}_i$ stands for the percentage of `people agents' that choose that same mobility mode once the iterative process has been performed. 

\begin{figure}[]
    \centering
    \includegraphics[scale = 0.42]{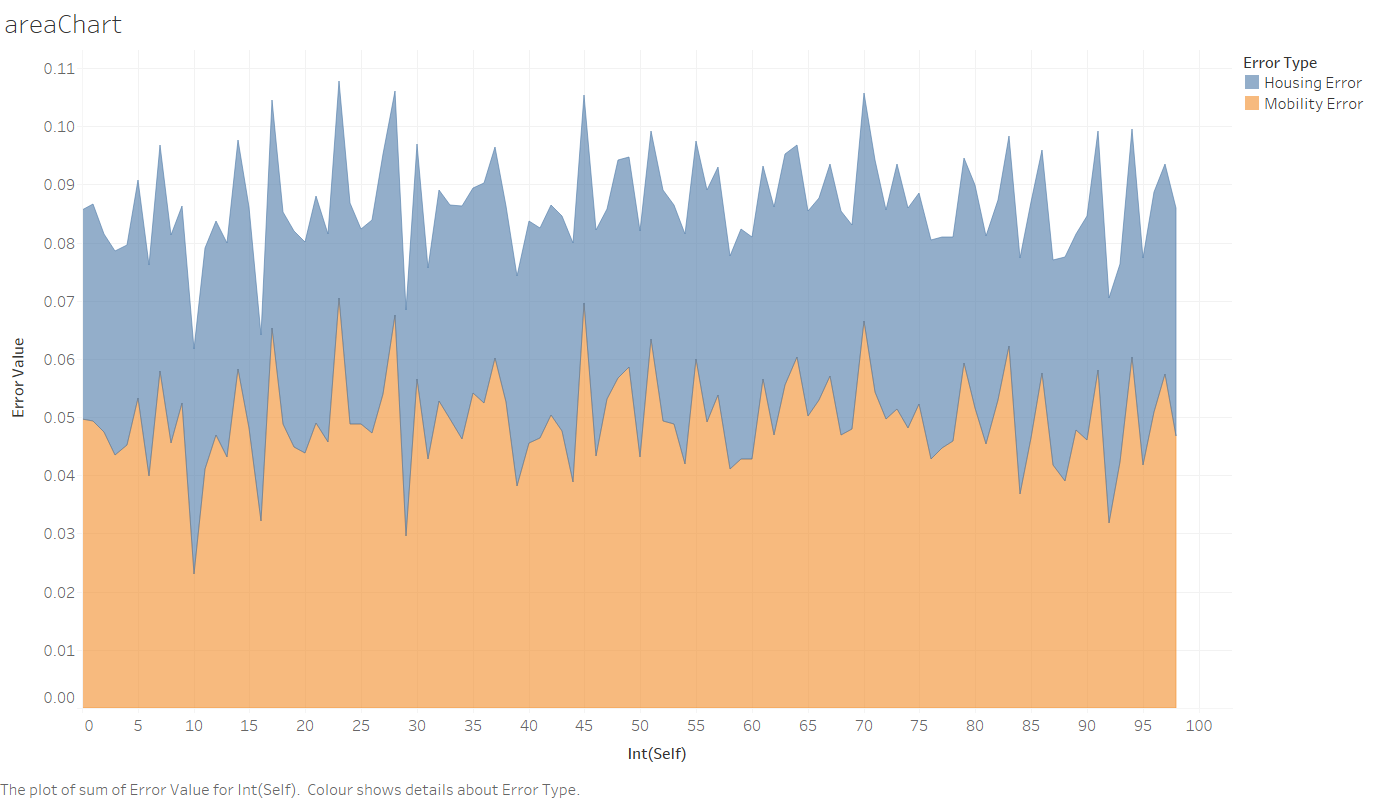}
    \caption{Evolution of housing and mobility errors [\%] in a hundred of the calibration process batch experiments.}
    \label{fig:areaChartError}
\end{figure}

As far as the exploration method is concerned, the hill climbing algorithm \cite{HillClimbing} has been applied and batch experiments have been performed, getting a reasonably good solution. The most suitable criteria combination has led to a mean housing error of 3.87\% and a mean mobility error of 2.30\% --Fig \ref{fig:areaChartError}--.\\

\section{Discussion} \label{discussion}

In the described model citizens' housing and mobility preferences are scored taking into account various parameters, which include price (for both housing and commuting), diversity acceptance, zone preference, time, difficulty of usage and social pattern. These parameters have been defined based on bibliography, considering that they should constitute a reduced set of variables that heavily affect the decision-making process. Should a finer-grained study be held, it should be noted that the zone preference criterion encompasses itself numerous possible attributes. The effect of greenery, high quality education centers or fresh food availability, to mention some, could also be included if a more detailed study is to be held. For a first version model, though, it has been assumed that the suggested approach was already complex enough. This same complexity, along with the metropolitan area scope, makes it challenging for the impacts resulting from changes in urban configurations to be calculated-real time, which represents a constraint when it comes to computational cost.
Finally, it should be noted that in order to calibrate the criteria in question, the exploration of the design space has been performed using the hill climbing algorithm. The minimization process has resulted in acceptable housing and mobility errors. However, this algorithm heavily relies on the selected starting point and, thus, alternative algorithms could be deployed if this limitation is to be avoided in future calibration processes.\\

\section{Conclusion} \label{conclusion}

This paper presents a generic methodology where citizens' criteria towards housing and mobility choices can be tested, as well as the resulting behavioural patterns monitored. Criteria regarding mobility and residential preferences have been calibrated and validated for the particular use case of Kendall Square, where major mobility issues have arisen as a consequence of the technological development of the surroundings and the explosion of housing prices. However, variations in input data and research into combinations of criteria that are particularised for a certain city are possible following the guidelines given earlier in the document. 
As part of the CityScope platform, this model enables the prediction of citizens' response to housing-related urban disruptions that aim to promote the pro-social development of cities. The usage of Agent-Based Modelling opens the door to the monitorisation of people's reactions to dynamically reconfigurable zoning policies. However, the deployment of such a data-driven platform requires of real-time feedback, so that different stakeholders can benefit from the instant impact analysis of the suggested actions. Further research regarding the conversion of this model into a real-time analysis tool is being held. This new approach will need to embrace the complexity of urban dynamics, while constituting an agile and dynamic decision-making tool.

\end{document}